\documentclass{PoS}

\usepackage{multirow}

\usepackage[utf8]{inputenc}

\usepackage{amsmath, amssymb}
\usepackage{mathtools}
\usepackage{ulem}

\def\slashchar#1{\setbox0=\hbox{$#1$}  % set a box for #1
   \dimen0=\wd0     % and get its size
   \setbox1=\hbox{/} \dimen1=\wd1  % get size of /
   \ifdim\dimen0>\dimen1   % #1 is bigger
      \rlap{\hbox to \dimen0{\hfil/\hfil}} % so center / in box
      #1     % and print #1
   \else     % / is bigger
      \rlap{\hbox to \dimen1{\hfil$#1$\hfil}} % so center #1
      /      % and print /
   \fi}

\newcommand{\dd}{\mathrm{d}}

\title{Chiral symmetry restoration by parity doubling and the structure of neutron stars}

\ShortTitle{}

\author{\speaker{Micha\l{} Marczenko}\\
        Institute of Theoretical Physics, University of Wroc\l{}aw, PL-50204 Wroc\l{}aw, Poland\\
        E-mail: \email{michal.marczenko@ift.uni.wroc.pl}
        }
\author{David Blaschke\\
        Institute of Theoretical Physics, University of Wroc\l{}aw, PL-50204 Wroc\l{}aw, Poland\\
        Bogoliubov Laboratory of Theoretical Physics, Joint Institute for Nuclear Research, 141980 Dubna, Russia\\
        National Research Nuclear University, 115409 Moscow, Russia
        }
\author{Krzysztof Redlich\\
        Institute of Theoretical Physics, University of Wroc\l{}aw, PL-50204 Wroc\l{}aw, Poland\\
        Extreme Matter Institute EMMI, GSI, D-64291 Darmstadt, Germany
        }
\author{Chihiro Sasaki\\
        Institute of Theoretical Physics, University of Wroc\l{}aw, PL-50204 Wroc\l{}aw, Poland\\
        }

\abstract{Recent lattice QCD studies at vanishing density exhibit the parity-doubling structure for the low-lying baryons around the chiral crossover temperature. This finding is likely an imprint of the chiral symmetry restoration in the baryonic sector of QCD, and is expected to occur also in cold dense matter, which makes it of major relevance for compact stars. By contrast, typical effective models for compact star matter embody chiral physics solely in the deconfined sector, with quarks as degrees of freedom. In this contribution, we present a description of QCD matter based on the effective hybrid quark-meson-nucleon model. Its characteristic feature is that, under neutron-star conditions, the chiral symmetry is restored in a first-order phase transition deep in the hadronic phase, before the deconfinement of quarks takes place. We discuss the implications of the parity doubling of baryons on the mass-radius relation for compact stars obtained in accordance with the modern constraints on the mass from PSR J0348+0432, the compactness from GW170817, as well as the direct URCA process threshold. We show that the existence of high-mass stars might not necessarily signal the deconfinement of quarks.}

\FullConference{XIII Quark Confinement and the Hadron Spectrum - Confinement2018\\
		31 July - 6 August 2018\\
		Maynooth University, Ireland}

\begin{document}

%%%%%%%%%%%%%%%%%%%%%%%%%%%%%%%%%%%%%%%%%%%%%%%%%%%%%%%%%%%%%%%
\section{Introduction}
\label{sec:introduction}
%%%%%%%%%%%%%%%%%%%%%%%%%%%%%%%%%%%%%%%%%%%%%%%%%%%%%%%%%%%%%%%
   
   The investigation of the equation of state~(EoS) of compact star matter has become rather topical within the past few years, mainly due to the one-to-one correspondence between EoS and \mbox{mass-radius}~(M-R) relationship~\cite{Lindblom:1998} for the corresponding sequence of compact stars via the solution of the Tolman-Oppenheimer-Volkoff (TOV) equations~\cite{Tolman:1939jz,Oppenheimer:1939ne}. Masses and radii of pulsars are targets of observational programs, which can therefore provide stringent constraints to the EoS and phase structure of quantum chromodynamics (QCD) in a region of the QCD phase diagram that is inaccessible to terrestrial experiments and present techniques of lattice QCD simulations. For the extraction of the compact star EoS via Bayesian analysis techniques using mass and radius measurements as priors see Refs.~\cite{Steiner:2010fz, Steiner:2012xt, Alvarez-Castillo:2016oln}. In the era of multimessenger astronomy, it shall soon become possible to constrain the sequence of stable compact star configurations in the \mbox{mass-radius} plane inasmuch that a benchmark for the EoS of cold and dense matter can be deduced from it.
   
   In the study of cold and dense QCD and its applications, commonly used are separate effective models for the nuclear and quark matter phases (two-phase approaches) with {\it a priori} assumed first-order phase transition, typically associated with simultaneous chiral and deconfinement transitions~\cite{Alvarez-Castillo:2016wqj}. If strong enough, such a strong phase transition manifests itself by the appearance of an almost horizontal branch on which the hybrid star solutions lie, as opposed to the merely vertical branch of pure neutron stars. In the literature, this strong phase transition has been discussed as due to quark deconfinement~\cite{Alvarez-Castillo:2016wqj}. This conclusion, as we demonstrate in this work, may however be premature since strong phase transitions with a large latent heat occur also within hadronic matter, e.g., due to the chiral restoration transition inside the nuclear matter phase, which we discuss here.
   
   In this contribution, we explore the implications of dynamical sequential phase transitions at high baryon density on the structure of neutron stars. To this end, we employ the hybrid quark-me\-son-nuc\-le\-on (QMN) model~\cite{Marczenko:2018jui}. We demonstrate how \mbox{high-mass} star configurations can be achieved with different neutron-star interior. We find that, depending on the parametrization, the model predicts different stable configurations with similar mass $M\simeq2~M_\odot$. Our main focus is on the role of the chiral symmetry restoration in the \mbox{high-mass} part of the \mbox{mass-radius} sequence.
   
   In the model parametrization, we pay special attention to recent observational constraints from neutron-star physics, in particular from the precise measurement of the high mass pulsar PSR~J0348+432~\cite{Antoniadis:2013pzd}, and for the compactness constraint from the GW170817 event~\cite{TheLIGOScientific:2017qsa}. Furthermore, we consider the so-called ``direct URCA (DU) constraint''.
   
   This paper is organized as follows. In Sec.~\ref{sec:hqmn_model}, we introduce the hybrid quark-meson-nucleon model. In Sec.~\ref{sec:eos}, we discuss the obtained numerical results on the equation of state under the neutron-star conditions. In Sec.~\ref{sec:results}, we discuss the obtained neutron-star relations, the direct URCA process, the tidal deformability, as well as possible realizations of the low-temperature phase diagram. Finally, Sec.~\ref{sec:conclusion} is devoted to conclusions.
   
%%%%%%%%%%%%%%%%%%%%%%%%%%%%%%%%%%%%%%%%%%%%%%%%%%%%%%%%%%%%%%%
\section{Hybrid Quark-Meson-Nucleon Model}
\label{sec:hqmn_model}
%%%%%%%%%%%%%%%%%%%%%%%%%%%%%%%%%%%%%%%%%%%%%%%%%%%%%%%%%%%%%%%

   In this section, we introduce the hybrid QMN model for the QCD phase transitions at finite temperature and density, following Ref.~\cite{Marczenko:2018jui}. In this contribution we consider an isospin-asymmetric system of two flavors, $N_f = 2$, in the mean-field approximation. Instead of the conventional Gel-Mann--Levy model, where the mass of the nucleon is generated through the non-vanishing sigma expectation value, the model considers another realization, where an introduction of a finite mass term does not break the chiral symmetry~\cite{Detar:1988kn,Jido:1999hd}. Hence, the sigma condensation generates only the mass difference between two chiral partners. The quark sector, on the other hand, is modeled in the standard linear-sigma model.
   
   In the mean-field approximation, the thermodynamic potential of the hybrid QMN model reads
   \begin{equation}\label{eq:thermo_pot}
      \Omega = \sum_{x}\Omega_x + V_\sigma + V_\omega + V_b + V_\rho\textrm,
   \end{equation}
   where the summation in the first term goes over the up~($u$) and down~($d$) quarks, as well as the nucleonic states with positive and negative parity. The positive-parity nucleons correspond to the positively charged and neutral $N(938)$ states, i.e., proton ($p_+$) and neutron ($n_+$), respectively. The \mbox{negative-parity} nucleons are identified as their counterparts, $N(1535)$~\cite{Patrignani:2016xqp}, and are denoted as $p_-$ and $n_-$. The kinetic term, $\Omega_x$, reads
   \begin{equation}
      \Omega_x = \gamma_x \int\frac{\dd^3p}{\left(2\pi\right)^3} T \left[\ln\left(1-n_x\right) + \ln\left(1-\bar n_x\right)\right]\textrm.
   \end{equation}
   The factor \mbox{$\gamma_\pm=2$} denotes the spin degeneracy for the nucleons with positive/negative parity, and \mbox{$\gamma_q=2\times 3 = 6$} is the spin-color degeneracy factor for the quarks. The functions $n_x$ are the modified \mbox{Fermi-Dirac} distribution functions for the nucleons
   \begin{equation}\label{eq:cutoff_nuc}
      n_\pm = \frac{1}{1+e^{\beta \left(E_\pm - \mu_\pm\right)}} \theta \left(\alpha^2 b^2 - \boldsymbol p^2\right) \textrm, \qquad \qquad \bar n_\pm = \frac{1}{1+e^{\beta \left(E_\pm + \mu_\pm\right)}} \theta \left(\alpha^2 b^2 - \boldsymbol p^2\right) \textrm,
   \end{equation}
   and for the quarks, accordingly
   \begin{equation}\label{eq:cutoff_quark}
      n_q = \frac{1}{1+e^{\beta \left(E_q - \mu_q\right)}} \theta \left(\boldsymbol p^2-b^2\right) \textrm, \qquad \qquad \bar n_q =  \frac{1}{1+e^{\beta \left(E_q + \mu_q\right)}} \theta \left(\boldsymbol p^2-b^2\right)\textrm,
   \end{equation}
   where $\beta$ is the inverse temperature, the dispersion relation $E_x = \sqrt{\boldsymbol p^2 + m_x^2}$. The effective chemical potentials for $p_\pm$ and $n_\pm$ are defined as\footnote{In the mean-field approximation, the nonvanishing expectation value of the $\omega$ field is the timelike component; hence we simply denote it by $\omega_0 \equiv \omega$. Similarly for the timelike and neutral component of the $\rho$ meson, $\rho_{03} \equiv \rho$.}
   \begin{equation}\label{eq:u_eff_had}
      \mu_{p_\pm} = \mu_B - g_\omega\omega - \frac{1}{2}g_\rho \rho + \mu_Q\textrm, \qquad\qquad \mu_{n_\pm} = \mu_B - g_\omega\omega + \frac{1}{2}g_\rho \rho\textrm.
   \end{equation}
   The constants $g_\omega$ and $g_\rho$ couple the nucleons to the $\omega$ and $\rho$ fields, respectively. The effective chemical potentials for up and down quarks are given by
   \begin{equation}\label{eq:u_effq}
      \mu_u = \frac{1}{3}\mu_B + \frac{2}{3}\mu_Q\textrm,\qquad \qquad \mu_d = \frac{1}{3}\mu_B - \frac{1}{3}\mu_Q\textrm.
   \end{equation}
   In Eqs.~(\ref{eq:u_eff_had})~and~(\ref{eq:u_effq}), $\mu_B$, $\mu_Q$ are the baryon and charge chemical potentials, respectively. In Eqs.~\eqref{eq:cutoff_nuc} and~\eqref{eq:cutoff_quark}, $b$ is the expectation value of the $b$-field, and $\alpha$ is a dimensionless model parameter~\cite{Benic:2015pia, Marczenko:2017huu, Marczenko:2018jui}. From the definition of $n_\pm$ and $n_q$, it is evident that, to mimic the statistical confinement, the expected behavior of the $b$ field is to have a nontrivial vacuum expectation value, in order to favor the hadronic degrees of freedom over the quark ones at low densities. On the other hand, it is expected that it vanishes at higher densities in order to suppress the hadronic degrees of freedom and to allow for the population of quarks. This is achieved by allowing $b$ to be generated from a potential $V_b$ (to be introduced later in this section).

   The effective masses of the parity doublers \mbox{$m_{p_\pm} = m_{n_\pm} \equiv m_\pm$} are given by
   \begin{equation}\label{eq:mass_had}
      m_\pm = \frac{1}{2}\left[\sqrt{\left(g_1+g_2\right)^2\sigma^2 + 4m_0^2} \mp \left( g_1 - g_2 \right) \sigma \right] \textrm,
   \end{equation}
   and for quarks, $m_u = m_d \equiv m_q$,
   \begin{equation}\label{eq:mass_quark}
      m_q = g_\sigma \sigma \textrm.
   \end{equation}
   The parameters $g_1$, $g_2$, $g_\sigma$ are Yukawa-coupling constants, $m_0$ is the chirally invariant mass of the baryons and is treated as an external parameter (see Ref.~\cite{Marczenko:2017huu}). The values of those couplings can be determined by fixing the fermion masses. We take the vacuum masses of the positive- and negative-parity hadronic states to be \mbox{$m_+=939~$MeV} and \mbox{$m_-=1500~$MeV}, respectively. The quark mass is assumed to be $m_+ = 3 m_q$ in the vacuum.
   
   The potentials in Eq.~(\ref{eq:thermo_pot}) are given as
   \begin{subequations}\label{eq:potentials}
   \begin{align}
      V_\sigma &= -\frac{\lambda_2}{2}\left(\sigma^2 + \boldsymbol\pi^2\right) + \frac{\lambda_4}{4}\left(\sigma^2 + \boldsymbol\pi^2\right)^2 - \epsilon\sigma \textrm,\\
      V_\omega &= -\frac{1}{2}m_\omega^2 \omega_\mu\omega^\mu\textrm,\\
      V_b &= -\frac{1}{2} \kappa_b^2 b^2 + \frac{1}{4}\lambda_b b^4 \textrm,\\
      V_\rho &= - \frac{1}{2}m_\rho^2{\boldsymbol \rho}_\mu{\boldsymbol \rho}^\mu \textrm,
   \end{align}
   \end{subequations}
   where $m_\omega$ and $m_\rho$ are the masses of the $\omega$ and $\rho$ mesons, respectively. The parameters $\lambda_2$, $\lambda_4$, and $\epsilon$ are
   \begin{equation}\label{eq:parity_params}
      \lambda_2 = \frac{m_\sigma^2 - 3 m_\pi^2}{2} \textrm{, }\;\;\;\; \lambda_4 = \frac{ m_\sigma^2 - m_\pi^2 }{2f_\pi^2}\textrm{, }\;\;\;\; \epsilon = m_\pi^2 f_\pi \textrm,
   \end{equation}
   where the pion mass $m_\pi=138~$MeV, and pion decay constant $f_\pi=93~$MeV. $m_\sigma$ is the mass of the $\sigma$ meson. In-medium profiles of the mean fields are obtained by extremizing the thermodynamic potential~(\ref{eq:thermo_pot}).
   
   In the grand canonical ensemble, the net-baryon number density for a species $x$ is defined as
   \begin{equation}
      \rho^x_B = - \frac{\partial \Omega_x}{\partial \mu_B}
   \end{equation}
   The total net-baryon number density is then the sum over all species, namely
   \begin{equation}
      \rho_B = \rho_B^{p_+} + \rho_B^{p_-} + \rho_B^{n_+} + \rho_B^{n_-} + \rho_B^{u} + \rho_B^{d} \textrm.
   \end{equation}
   The particle-density fractions are defined as 
   \begin{equation}
      Y_x = \frac{\rho_B^x}{\rho_B} \textrm.
   \end{equation}
   
   Following the previous studies, we study $m_0=790$~MeV~\cite{Zschiesche:2006zj, Benic:2015pia, Marczenko:2018jui}. We take four representative values of the $\alpha$ parameter, namely $\alpha b_0 = 350,~370,~400,~450~$MeV, in order to systematically study how they influence the three constraints, i.e., the $2~M_\odot$, the tidal deformability, and the direct URCA constraint. The model parameters are fixed as in Ref.~\cite{Marczenko:2018jui}.
   
   In the next section we discuss the influence of the external parameter $\alpha$ on the equation of state in the hybrid QMN model and its impact on the chiral phase transition, under the neutron-star conditions of $\beta$ equilibrium and charge neutrality.
   
%%%%%%%%%%%%%%%%%%%%%%%%%%%%%%%%%%%%%%%%%%%%%%%%%%%%%%%%%%%%%%%
\section{Equation of State}
\label{sec:eos}
%%%%%%%%%%%%%%%%%%%%%%%%%%%%%%%%%%%%%%%%%%%%%%%%%%%%%%%%%%%%%%%
   
   \begin{figure}[t!]
   \begin{center}
      \includegraphics[width=.5\linewidth]{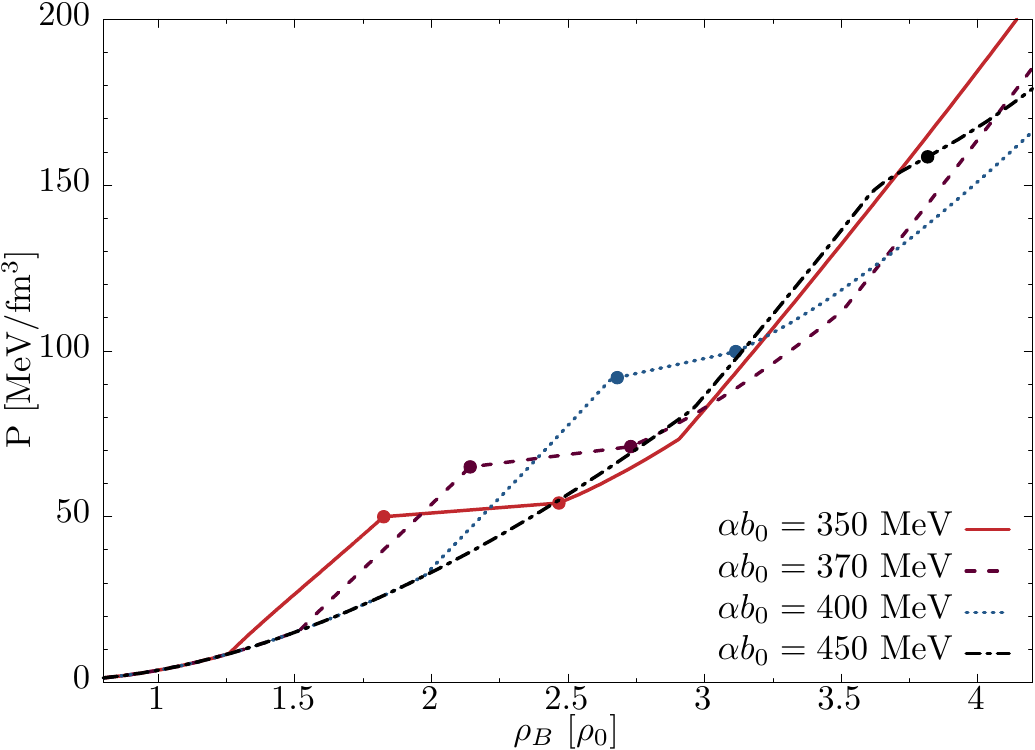}
      \caption{Thermodynamic pressure $P$ as a function of the net-baryon number density $\rho_B$, in units of the saturation density, $\rho_0=0.16$~fm$^{-3}$ for $m_0=790~$MeV. The regions between circles correspond to the coexistence of chirally broken and restored phases in the first-order phase transition. For \mbox{$\alpha b_0=450~$MeV} the transition is a crossover. The deconfinement transitions are triggered at higher densities and are not shown here.}
      \label{fig:eos}
   \end{center}
   \end{figure}
   
   The composition of neutron-star matter requires \mbox{$\beta$ equilibrium}, as well as the charge neutrality condition. To this end, we include electrons and muons as gases of free relativistic particles. In Fig.~\ref{fig:eos}, we show the calculated zero-temperature equations of state in the mean-field approximation with $m_0=790~$MeV, for different values of the $\alpha$ parameter. The mixed phases of the chirally broken and restored phases are shown between circles. We stress that the chiral and hadron-to-quark phase transitions are sequential. The latter happen at higher densities and are not shown in the figure. In all cases, the \mbox{low-density} behavior is similar. In the case of $\alpha b_0=350$~MeV, the chiral phase transition is triggered at roughly $1.82~\rho_0$, with the mixed phase persisting up to $2.46~\rho_0$. Higher values of the $\alpha$ parameter yield weaker transitions at higher densities. For $\alpha b_0=370$~MeV, the mixed phase appears between $2.14$ and $2.73~\rho_0$, and for $\alpha b_0=400$~MeV between $2.68$ and $3.11~\rho_0$. On the other hand, for $\alpha b_0=450$~MeV, the transition turns into a crossover at roughly $3.81~\rho_0$. This stays in correspondence to the case of isospin-symmetric matter, where higher value of $\alpha$ weakens the first-order chiral phase transition, which goes through a critical point, and eventually turns into a crossover transition~\cite{Marczenko:2017huu}.
   
   \begin{figure}[t!]
   \begin{center}
         \includegraphics[width=0.49\linewidth]{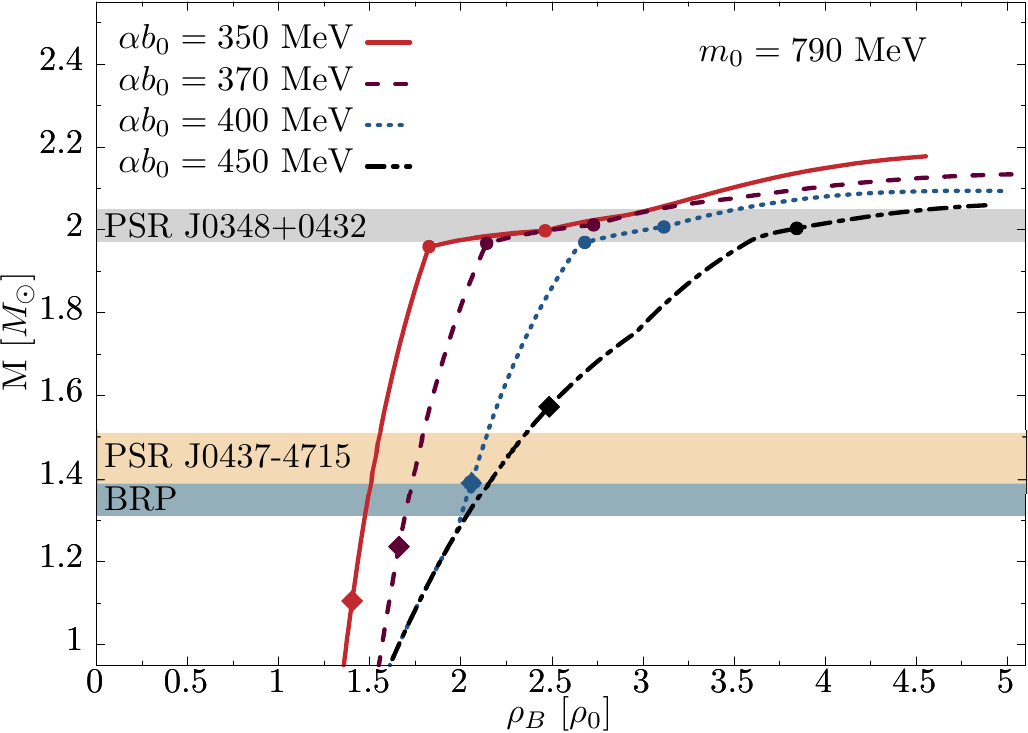}
         \includegraphics[width=0.49\linewidth]{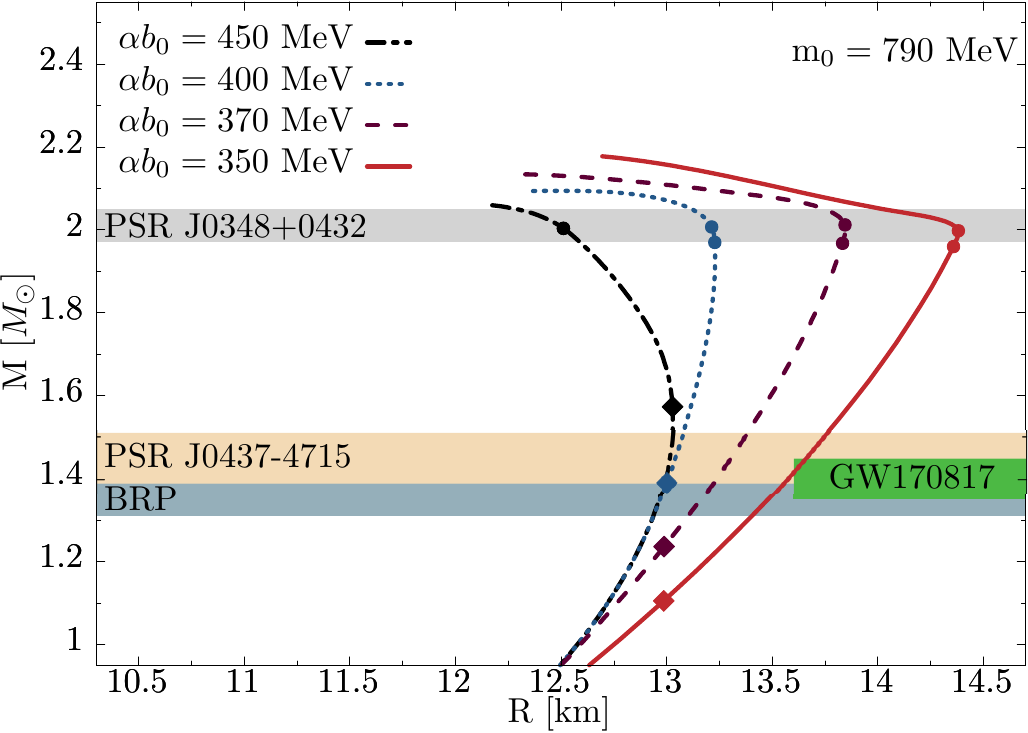}
      \caption{Sequences of masses for compact stars vs their central net-baryon density (left panel) and vs radius (right panel) as solutions of the TOV equations for $m_0=790~$MeV and four different cases of \mbox{$\alpha b_0=350,~370,~400,~450$~MeV}. The regions between the circles show the coexistence of the chirally broken and chirally restored phases. The diamonds indicate the threshold mass for the direct URCA process (see Sec.~\ref{sec:direct_urca}). The upper gray band is the $2.01(4)~M_\odot$ observational constraint~\cite{Antoniadis:2013pzd}; the orange band, $1.44(7)~M_\odot$, shows the mass of PSR~J0437-4715 for which the NICER experiment~\cite{Reardon:2015kba} will soon provide a radius measurement~\cite{Miller:2016pom}. Finally, the lower blue band is the $1.35(4)~M_\odot$ binary radio pulsar (BRP) constraint~\cite{Thorsett:1998uc}. The green areas (GW170817) in the right panel show excluded radii of $1.4~M_\odot$ stars extracted in Ref.~\cite{Annala:2017llu}.}
      \label{fig:m_density}
   \end{center}
   \end{figure}
   
%%%%%%%%%%%%%%%%%%%%%%%%%%%%%%%%%%%%%%%%%%%%%%%%%%%%%%%%%%%%%%%
\section{Results}
\label{sec:results}
%%%%%%%%%%%%%%%%%%%%%%%%%%%%%%%%%%%%%%%%%%%%%%%%%%%%%%%%%%%%%%%
   
%%%%%%%%%%%%%%%%%%%%%%%%%%%%%%%%%%%%%%%%%%%%%%%%%%%%%%%%%%%%%%%
   \subsection{TOV solutions for compact star sequences}
   \label{sec:mass_radius}
%%%%%%%%%%%%%%%%%%%%%%%%%%%%%%%%%%%%%%%%%%%%%%%%%%%%%%%%%%%%%%%
      
      We use the equations of state introduced in the previous section (see Fig.~\ref{fig:eos}) to solve the general-relativistic TOV equations~\cite{Tolman:1939jz, Oppenheimer:1939ne} for spherically symmetric objects. In the left panel of Fig.~\ref{fig:m_density}, we show the relationship of mass versus central net-baryon density, for the calculated sequences of compact stars, together with the state-of-the-art constraint on the maximum mass for the pulsar PSR~J0348+0432~\cite{Antoniadis:2013pzd}. Corresponding mass-radius relations are shown in the right panel of Fig.~\ref{fig:m_density}. We point out that the chiral restoration transition leads to a softening of the EoS so that it is accompanied by a strong increase of the central densities, while the mass of the star is almost unchanged.
      
      In general, there is one-to-one correspondence between an EoS and the \mbox{mass-radius} relation calculated with it. The three curves for \mbox{$\alpha b_0 = 350,~370,~400~$MeV} consist of three phases; the chirally broken phase in the low-mass part of the sequence, the chirally restored phase in the \mbox{high-mass} part, and the mixed phase between filled circles. Similarly to the equation of state, increasing the value of $\alpha$ softens the chiral transition, which eventually becomes a smooth crossover for $\alpha b_0 = 450~$MeV and consists only of branches with chiral symmetry being broken and restored, separated by a circle.
      
      We note that the end points of the lines correspond to the onset of quark degrees of freedom in each case, after which the equation of state is not stiff enough to sustain the gravitational collapse and the branches become immediately unstable. This is because, in the current model setup, quarks are not coupled with the vector field leading to a repulsive force. On the other hand, it is known that repulsive interactions tend to stiffen the equation of state. Hence, an additional repulsive force in the quark sector could possibly make the branch stiff enough, so that an additional family of stable hybrid compact stars would appear, with the possibility for the \mbox{high-mass} twin scenario advocated by other effective models~\cite{Alvarez-Castillo:2017qki, Ayriyan:2017nby, Kaltenborn:2017hus}. We leave a further study of quark matter as our future work.
      
      \begin{figure}[t!]
      \begin{center}
         \includegraphics[width=.49\linewidth]{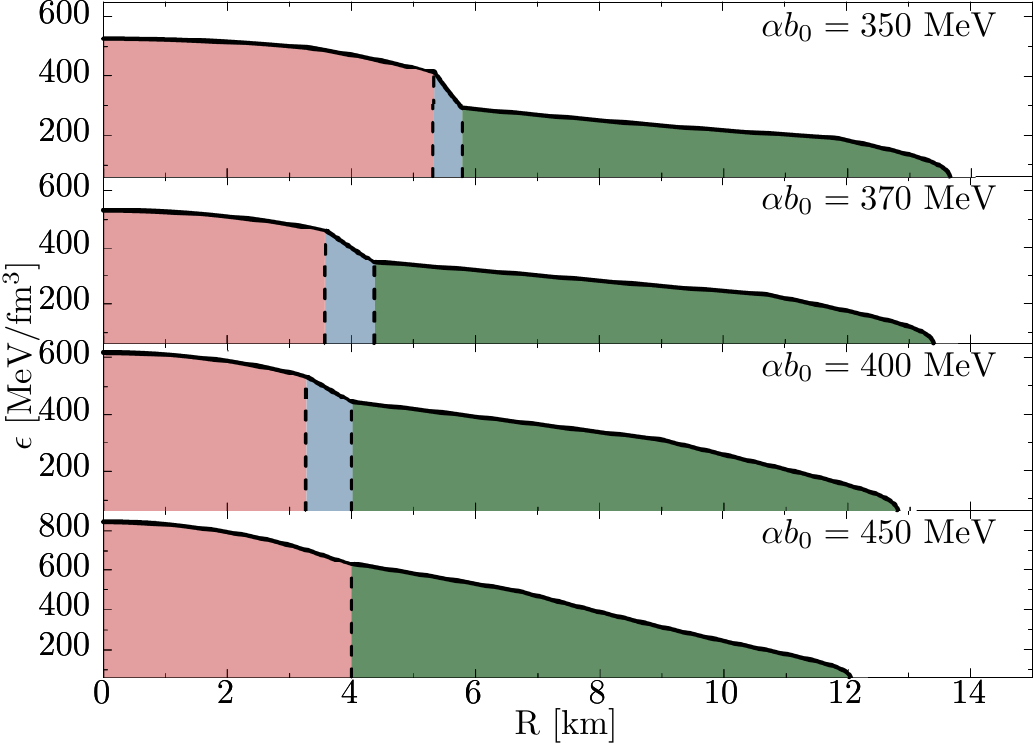}
         \includegraphics[width=.49\linewidth]{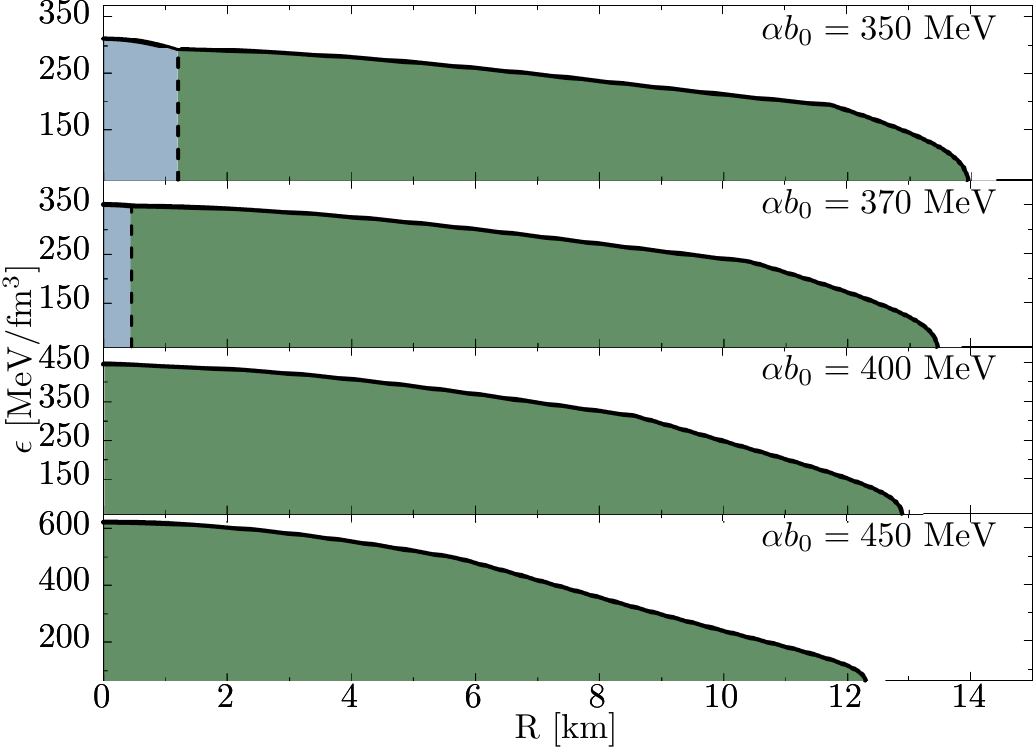}
         \caption{Profiles of the energy density for neutron stars with \mbox{$M = 1.97~M_\odot$} (right panel) and \mbox{$M = 2.05~M_\odot$} (left panel) for \mbox{$m_0=790~$MeV}, for four different cases \mbox{$\alpha b_0=350,~370,~400,~450$~MeV}. The green regions show the phase, where the chiral symmetry is broken; in the red regions chiral symmetry is restored, whereas the blue regions indicate the regions of coexistence of both phases (mixed phase).}
         \label{fig:profiles}
      \end{center}
      \end{figure}
      
      Notably, the chiral transition for all values of $\alpha b_0$ occurs in the \mbox{high-mass} part of the sequence, namely in the vicinity of the \mbox{$2~M_\odot$} constraint, followed by a rapid flattening of the \mbox{mass-radius} sequence. The transitions are, however, not strong enough to produce an additional family of solutions, disconnected by an unstable branch.
      
      We note that the obtained \mbox{mass-radius} relations stay in good agreement with the low-mass constraints derived from the recent neutron-star merger, namely that the radius of a $1.6~M_\odot$ neutron star must be larger than $10.68^{+0.15}_{-0.04}$~km~\cite{Bauswein:2017vtn}. Also, the recently extracted constraint on the radius of $1.4~M_\odot$ stars, that is \mbox{$12.0$~km$~<R(1.4~M_\odot) < 13.6~$km~\cite{TheLIGOScientific:2017qsa, Annala:2017llu, Tews:2018iwm, Most:2018hfd}}, is well fulfilled by all of the parametrizations. In Fig.~\ref{fig:m_density}, the excluded upper bound of this constraint is shown as the green region. Since the radii at $1.4~M_\odot$ are well above $12.0~$km, we do not plot the lower bound of the constraint for the sake of clarity of the figure. From this, one sees that values of $\alpha b_0$ lower than roughly $350~$MeV are excluded by the maximal radius constraint.
      
      In Fig.~\ref{fig:profiles}, we show the energy density profiles of stars for $m_0=790~$MeV. The bottom panel shows profiles of stars with mass $M = 1.97~M_\odot$. The cases with $\alpha b_0=400$ and $450~$MeV show the profiles, the interiors of which consist only of nuclear matter with broken chiral symmetry (green region). For the stars with $\alpha b_0=350$ and $370~$MeV, the chiral transitions are triggered at roughly $1.2$ and $0.4$~km radii, respectively, and in their cores only the mixed phase (blue region) is realized. The radii of these stars vary from $13.7$ to $14.3$~km. In the top panel, we show profiles of stars with mass $M=2.05~M_\odot$. The chiral transitions are triggered in all four cases, and the chirally restored phase is reached in their cores (red region). For $\alpha b_0=350~$MeV, the mixed phase is featured between $5.3$ and $5.8$~km radius, for $\alpha b_0=370~$MeV between $3.6$ and $4.4$~km, and for $\alpha b_0=400~$MeV between $3.3$ and $4.1$~km. For $\alpha b_0=450~$MeV, as discussed in previous sections, the transition is a smooth crossover. In this case, we assume the chirally broken and chirally restored phases are separated by the peak in $\partial \sigma / \partial \mu_B$, which happens around $4.0~$km radius. The radii of these stars vary from $12.2$ to $14.0$~km.
      
%%%%%%%%%%%%%%%%%%%%%%%%%%%%%%%%%%%%%%%%%%%%%%%%%%%%%%%%%%%%%%%
   \subsection{Direct URCA process}
   \label{sec:direct_urca}
%%%%%%%%%%%%%%%%%%%%%%%%%%%%%%%%%%%%%%%%%%%%%%%%%%%%%%%%%%%%%%%
      
      The DU process, $n_+ \rightarrow p_+ + e + \bar \nu_e$, is essential for the cooling of neutron stars and is not expected to occur in neutron stars with masses of the order of $1-1.5~M_\odot$~\cite{Klahn:2006ir}. When triggered, it leads to a substantial enhancement of the neutrino emission, and hence to the neutron-star cooling rates. The DU process becomes operative when a critical value of the proton fraction is exceeded~\cite{Lattimer:1991ib}. Taking into account the presence of the parity doublers, as well as assuming that below the deconfinement transition quarks are not populated, i.e. $\rho_B^u=\rho_B^d=0$, the proton fraction is given by
      \begin{equation}\label{eq:p_frac}
         Y_{p_+} = \frac{\rho^{p_+}_B}{\rho_B} = \frac{\rho^{p_+}_B}{\rho^{p_+}_B+\rho^{p_-}_B+\rho^{n_+}_B+\rho^{n_-}_B}\textrm.
      \end{equation}
      
      The critical values of masses for the DU process are marked on the \mbox{mass-radius} profiles in Fig.~\ref{fig:m_density} as diamonds. Also shown is the binary-radio-pulsar mass region \mbox{$M=1.35(4)~M_\odot$}~\cite{Thorsett:1998uc} (lower blue band), which sets a lower-bound constraint for the DU threshold in neutron stars. From the figure, it is clear that the DU process becomes operative still in the chirally broken phase.
      
      Before the chiral transition takes place, the parity partners of proton and neutron are not populated; i.e., their densities are 0. The only relevant degrees of freedom are the positive-parity ground state nucleons and leptons. Hence, in the chirally broken phase, the situation is similar to the case of ordinary nuclear matter. In that case, the critical value for the proton fraction can be deduced~\cite{Klahn:2006ir},
      \begin{equation}\label{eq:yp1}
         Y^{\rm DU}_{p_+} = \frac{1}{1 + \left(1 + \sqrt[3]{Y_e} \right)^3} \textrm,
      \end{equation}
      where $Y_e = \rho_e / (\rho_e + \rho_\mu)$. The $\rho_e$ and $\rho_\mu$ are the electron and muon densities. $Y_e$ may vary from $1/2$ ($\rho_e = \rho_\mu$) to 1 ($\rho_\mu = 0$). For $Y_e=1/2$, the critical value is $14.8\%$, and it goes down to $11.1\%$ for the muon-free case.
      
      The above estimate changes when the negative-parity chiral partners are populated. To see this, let us assume a phase with fully restored chiral symmetry. In this limit, we expect that the parity doublers are degenerate; hence their densities are equal. The charge neutrality condition becomes $2\rho_{p_+} = \rho_e + \rho_\mu$, while the momentum conservation condition remains the same~\cite{Lattimer:1991ib}, because only the positive-parity states take part in the DU process. This might be modified due to the fact that neutrinos have a finite rest mass which results in a mixing of the left- and right-handed neutrino sectors. In the present work we do not elaborate on this interesting beyond-the-standard-model aspect, for which to the best of our knowledge at present no investigation exists in the literature. With this, one finds the threshold to be equal,
      \begin{equation}\label{eq:yp2}
         Y^{\rm DU}_{p_+} = \frac{1}{1 + \left(1 + \sqrt[3]{2Y_e} \right)^3} \textrm.
      \end{equation}
      For $Y_e=1/2$, the critical value is $11.1\%$, and it goes down to $8.0\%$ for the muon-free case. Note that this estimate is systematically lower than the one from Eq.~(\ref{eq:yp1}).
      
      We find that the critical proton fraction is reached already in the chirally broken phase, for all parametrizations. However, while the direct URCA process is always operative in the chirally restored phase at high densities (and thus high star masses), for the case $\alpha b_0=350$~MeV it becomes operative already in the chirally broken phase at rather low densities in stars of the typical mass range, so that this parametrization becomes highly unfavorable.
      
      \begin{figure}[t!]
      \begin{center}
         \includegraphics[width=.49\linewidth]{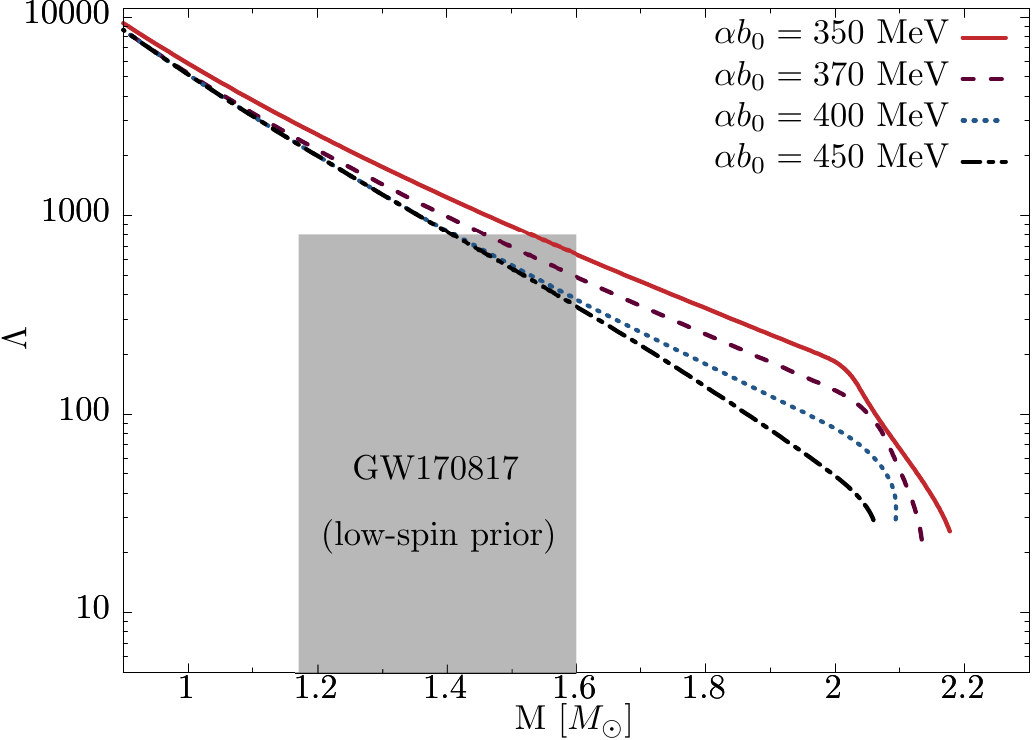}
         \includegraphics[width=.49\linewidth]{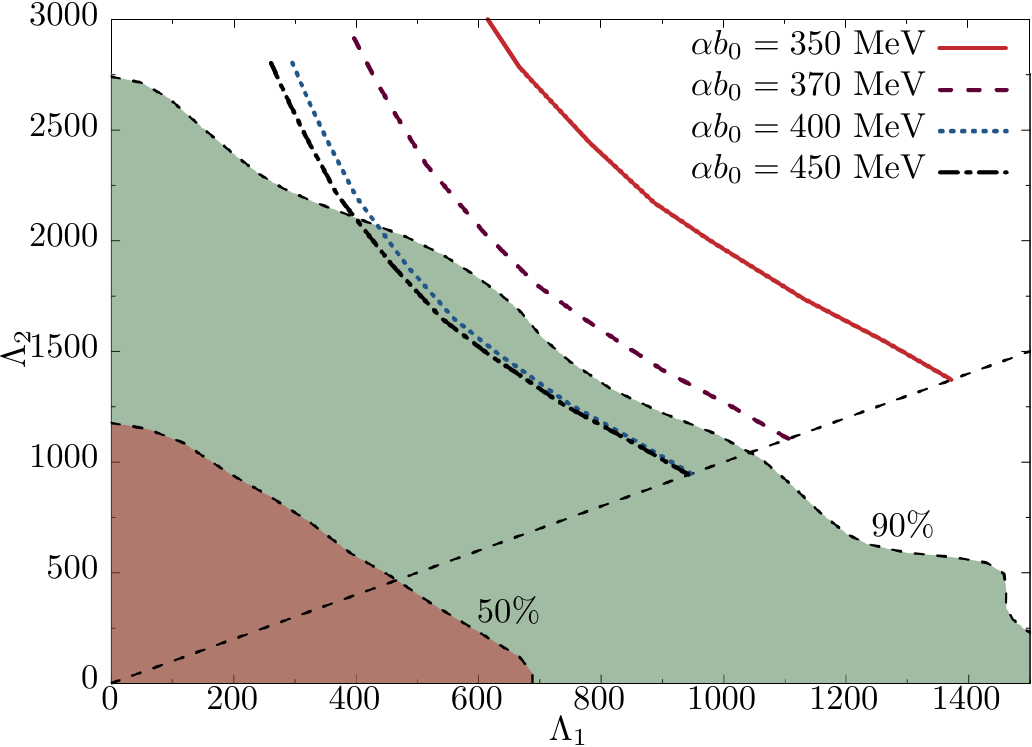}
         \caption{Left: Tidal deformability parameter $\Lambda$ as a function of neutron-star mass, for $m_0=790~$MeV. The gray box shows the $\Lambda<800$ constraint in the range $1.16-1.60~M_\odot$ of the low-spin prior~\cite{TheLIGOScientific:2017qsa}. Right: Corresponding tidal deformability parameters $\Lambda_1$ and $\Lambda_2$ of the low- and \mbox{high-mass} mergers obtained from the $\Lambda(m)$. Shown are also the $50\%$ and $90\%$ probability contours for the low-spin prior~\cite{TheLIGOScientific:2017qsa}.}
         \label{fig:tidal1}
      \end{center}
      \end{figure}
      
      While Eq.~(\ref{eq:yp1}) provides a very useful estimate for the DU threshold in the chirally broken phase, the estimate given by Eq.~(\ref{eq:yp2}) should be taken with caution. In the current approach, with two sequential phase transitions, the densities, where the parity doublers are fully degenerate, might be beyond the deconfinement transition. Hence, such a scenario would not occur. The situation is similar if the two transitions are simultaneous. In such case, the chiral symmetry restoration is accompanied by the transition from hadrons to quarks.
      
%%%%%%%%%%%%%%%%%%%%%%%%%%%%%%%%%%%%%%%%%%%%%%%%%%%%%%%%%%%%%%%
   \subsection{Tidal deformability}
   \label{sec:tidal}
%%%%%%%%%%%%%%%%%%%%%%%%%%%%%%%%%%%%%%%%%%%%%%%%%%%%%%%%%%%%%%%
      
      The dimensionless tidal deformability parameter $\Lambda$ can be computed through its relation to the Love number $k_2$~\cite{Hinderer:2007mb,Damour:2009vw,Binnington:2009bb,Yagi:2013awa,Hinderer:2009ca},
      \begin{equation}
         \Lambda = \frac{2}{3} k_2 C^{-5} \textrm,
      \end{equation}
      where $C = M/R$ is the star compactness parameter, with $M$, $R$ being the total mass and radius of a star. In the left panel of Fig.~\ref{fig:tidal1}, we show the dimensionless tidal deformability parameter $\Lambda$ as a function of neutron star mass $M$, for \mbox{$m_0=790~$MeV}. We also show the constraint derived in~\cite{TheLIGOScientific:2017qsa}, $\Lambda\left(1.4~M_\odot\right) < 800$. The constraint is met only for the cases with $\alpha b_0 = 400$ and $450~$MeV. In the right panel of Fig.~\ref{fig:tidal1}, we plot the tidal deformability parameters $\Lambda_1$ vs $\Lambda_2$ of the high- and low-mass members of the binary merger together with the $50\%$ and $90\%$ fidelity regions obtained by the LVC analysis of the GW170817 merger event~\cite{TheLIGOScientific:2017qsa}. We note that the tidal deformability parameter favors equations of state that are soft around the saturation density. On the other hand, the $2~M_\odot$ constraint requires a sufficiently stiff equation of state at higher densities. The interplay between the two constraints can be further used to fix the model parameters $m_0$ and $\alpha b_0$. We discuss this matter in the next subsection.
      
%%%%%%%%%%%%%%%%%%%%%%%%%%%%%%%%%%%%%%%%%%%%%%%%%%%%%%%%%%%%%%%
   \subsection{Isospin-symmetric phase diagram}
   \label{sec:qcd_phase_diagram}
%%%%%%%%%%%%%%%%%%%%%%%%%%%%%%%%%%%%%%%%%%%%%%%%%%%%%%%%%%%%%%%
      
      \begin{figure}[t!]
      \begin{center}
         \includegraphics[width=.5\linewidth]{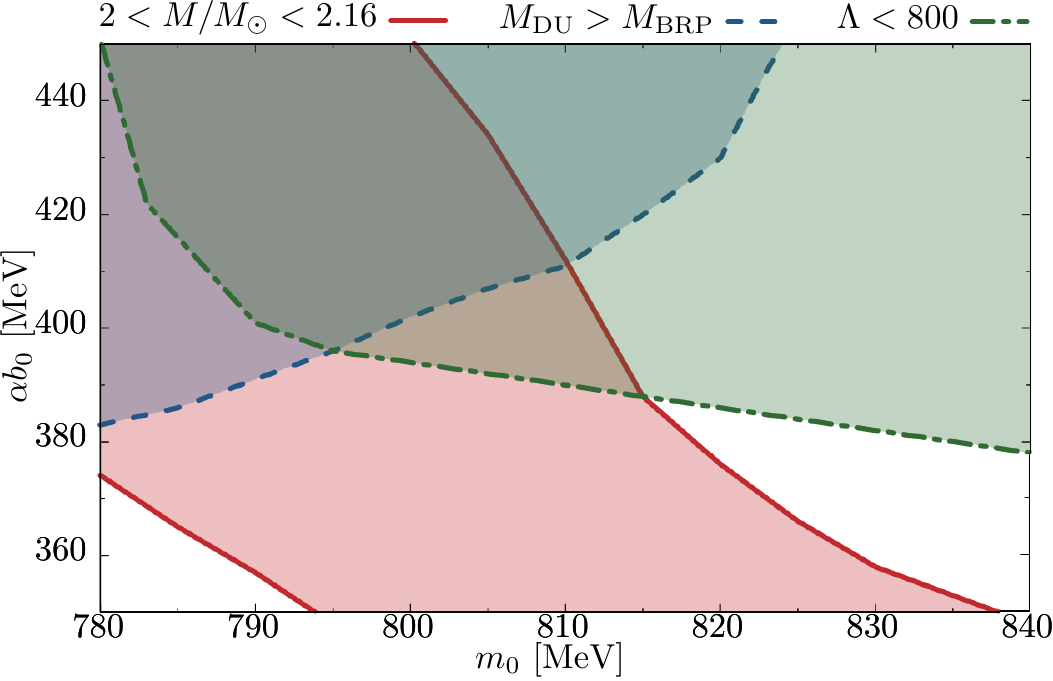}
         \includegraphics[width=.49\linewidth]{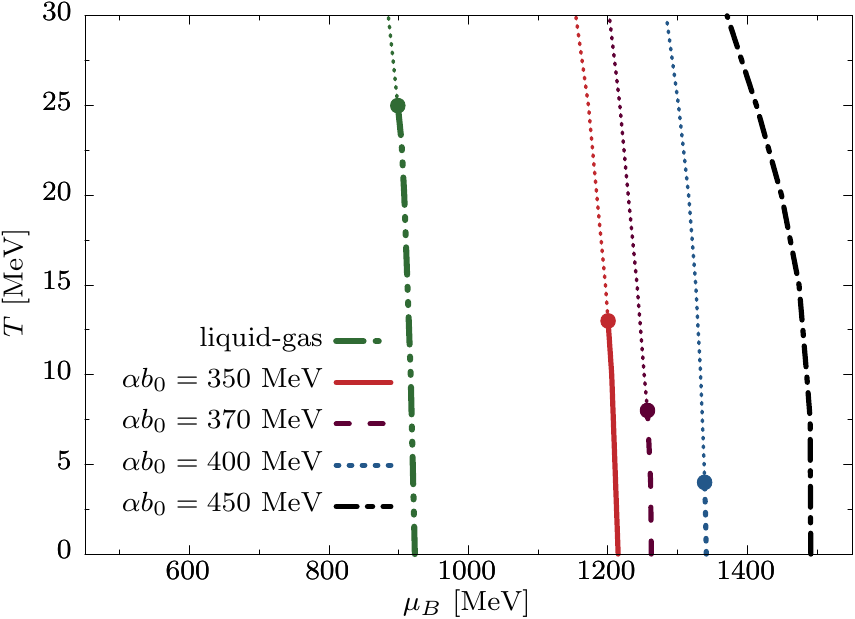}
         \caption{Left: Constraints for the model parameters $\alpha b_0$ and $m_0$: The maximum mass constraint (red, solid line), the direct URCA (blue, dashed line), and the tidal deformability constraint (green, broken-dashed line). The corresponding shaded areas show regions where the constraints are met. Right: Low-temperature part of the phase diagram in the $(T,\mu_B)$-plane for isospin-symmetric matter obtained in the hybrid QMN model. The green, \mbox{dashed-doubly-dotted} curve corresponds to the liquid-gas phase transition common for all $\alpha b_0$. The circles indicate critical points on the transition lines above which the first-order transition turns into a crossover. No critical point is shown for the case with $\alpha b_0=450$~MeV, for which the chiral transition is a smooth crossover at all temperatures.}
         \label{fig:constraints}
      \end{center}
      \end{figure}
      
      In the left panel of Fig.~\ref{fig:constraints}, we compile the three constraints discussed in this work in the $(\alpha b_0,m_0)$-plane. First, we show the constraint on the maximum mass of nonrotating, cold neutron stars~\cite{Rezzolla:2017aly,Most:2018hfd, Margalit:2017dij, Shibata:2017xdx} (red solid line), namely
      \begin{equation}
         2.01 < M/M_\odot < 2.16 \textrm.
      \end{equation}
      Secondly, we show the constraint on our parameter space that results from the direct URCA constraint, i.e., from the existence of a lower bound on the mass of neutron stars for which the fast direct URCA process could become operative. We take $M_{\rm BRP} \approx 1.4~M_\odot$, representing the upper limit for the masses of typical neutron stars. Finally, the constraint  on the tidal deformability parameter, $\Lambda(1.4~M_\odot) < 800$, induced from the GW170817 merger event analysis~\cite{TheLIGOScientific:2017qsa} is shown as the green broken-dashed line. The areas with corresponding shaded colors show regions where the constraints are met. The region where they overlap gives the most preferable sets of the external parameters $\alpha b_0$ and $m_0$. Since higher $m_0$, in general, yields a softer equation of state, the stiffness required by the maximum mass constraint is compensated by lower values of $\alpha b_0$, and inversely for the remaining two constraints. At lower values of $m_0$, the maximum mass constraint is met practically for the whole range of $\alpha b_0$, and the region of overlap is set by the other two constraints (top left corner of the figure). Hence, in general higher values of $\alpha b_0$ are more preferable. The three constraints are met down to roughly $m_0=780~$MeV. This sets the lower and upper bounds for the $m_0$ parameter. As a result, the allowed range for $m_0$ is roughly $m_0=780-810~$MeV.
      
      In the right panel of Fig.~\ref{fig:constraints}, we show the phase diagram obtained in the model in the $(T,\mu_B)$-pla\-ne, for the case of $m_0=790~$MeV. The first-order liquid-gas phase transition (green dashed-dou\-bly-dot\-ted line) develops a critical point around $T=25~$MeV, and turns into a crossover above it, similarly to the pure parity doublet model. We note that by construction of the hybrid QMN model, the liquid-gas phase transition line is common for all values of the $\alpha$ parameter~\cite{Benic:2015pia, Marczenko:2017huu}. Similar phase structure is seen in the chiral phase transition for $\alpha b_0=350,~370,~400~$MeV, which develop critical points at $T=13,~8,~4~$MeV, respectively. On the other hand, for $\alpha b_0 = 450~$MeV, the chiral transition is a smooth crossover at all temperatures. Note that, in view of the constraints discussed in this work (see Sec.~\ref{sec:results}), the scenarios with $\alpha b_0=350~$MeV and $\alpha b_0=370~$MeV are rather excluded. Hence, either rather low temperature for the critical end point or even its absence in the phase diagram for isospin-symmetric matter is favored.
      
%%%%%%%%%%%%%%%%%%%%%%%%%%%%%%%%%%%%%%%%%%%%%%%%%%%%%%%%%%%%%%%
\section{Conclusions}
\label{sec:conclusion}
%%%%%%%%%%%%%%%%%%%%%%%%%%%%%%%%%%%%%%%%%%%%%%%%%%%%%%%%%%%%%%%
   
   We investigated the consequences of a recently developed hybrid QMN model for the equation of state of dense matter under neutron-star conditions and the phenomenology of compact stars. We focused on the implications of the realization of the chiral symmetry restoration by parity doubling within the hadronic phase. We have demonstrated that a strong first-order phase transition invalidates the implication that a flattening, eventually even occurrence of a mass-twin phenomenon, of the \mbox{mass-radius} relation for compact stars at $2~M_\odot$, could inevitably signal a deconfinement phase transition in compact stars~\cite{Alvarez-Castillo:2016wqj}. An abrupt change in a \mbox{mass-radius} profile in the \mbox{high-mass} part of the sequence is, in general, a result of a phase transition. As we have discussed, in a model with two sequential transitions, it does not need to be associated with the deconfinement transition, and hence does not imply the existence of quark matter in the core of a neutron star.
   
   We have discussed how modern astrophysical constraints compiled together allow for better determination of the available parameter range, and demonstrated that not only too soft (excluded by the maximum mass constraint), but also too stiff (excluded by either the direct URCA or the tidal deformability constraint) equations of state may be ruled out in the current approach. Finally, we have shown that, due to the parity doubling phenomenon, the obtained results suggest rather low value of the temperature for the critical end point of the \mbox{first-order} chiral phase transition in the phase diagram, which eventually may even be absent.
   
   It would be of great interest to establish hybrid QMN equations of state that include hyperon degrees of freedom. In general the inclusion of heavier flavors is known to soften the equation of state and additional repulsive forces are needed to comply with the $2~M_\odot$ constraint. Also, additional stiffness from the quark side would play a role, which is not included in the current study. We note that it may be essential for establishing the branch of stable hybrid star solutions. Work in this direction is in progress and the results will be reported elsewhere.
   
%%%%%%%%%%%%%%%%%%%%%%%%%%%%%%%%%%%%%%%%%%%%%%%%%%%%%%%%%%%%%%%
\acknowledgments
%%%%%%%%%%%%%%%%%%%%%%%%%%%%%%%%%%%%%%%%%%%%%%%%%%%%%%%%%%%%%%%
   
   This work was partly supported by the Polish National Science Center (NCN), under Maestro Grant No. DEC-2013/10/A/ST2/00106 (K.R., C.S.), Opus Grant No. UMO-2014/13/B/ST9/02621 (M.M. and D.B.), and Preludium Grant No. UMO-2017/27/N/ST2/01973 (M.M.). D.B. is grateful for support within the MEPhI Academic Excellence program under Contract No. 02.a03.21.0005. We acknowledge the COST Actions CA15213 ``THOR'' and CA16214 ``PHAROS'' for supporting networking activities.

\end{document}